\documentclass[twocolumn,showpacs,amsmath,amssymb,10pt]{revtex4}
\usepackage{amsfonts}
\usepackage{bm}

\newcommand{\ot}{{\,\otimes\,}}
\newcommand{{\Cd}}{{\mathbb{C}^d}}
\newcommand{\sbsigma}{{\mbox{\scriptsize \boldmath $\sigma$}}}
\newcommand{\sbalpha}{{\mbox{\scriptsize \boldmath $\alpha$}}}
\newcommand{\sbbeta}{{\mbox{\scriptsize \boldmath $\beta$}}}
\newcommand{\bsigma}{{\mbox{ \boldmath $\sigma$}}}
\newcommand{\balpha}{{\mbox{ \boldmath $\alpha$}}}
\newcommand{\bbeta}{{\mbox{ \boldmath $\beta$}}}
\newcommand{\bmu}{{\mbox{ \boldmath $\mu$}}}
\newcommand{\bnu}{{\mbox{ \boldmath $\nu$}}}
\newcommand{\sbmu}{{\mbox{\scriptsize \boldmath $\mu$}}}
\newcommand{\sbnu}{{\mbox{\scriptsize \boldmath $\nu$}}}
\def\oper{{\mathchoice{\rm 1\mskip-4mu l}{\rm 1\mskip-4mu l}%
{\rm 1\mskip-4.5mu l}{\rm 1\mskip-5mu l}}}
\def\<{\langle}
\def\>{\rangle}

\begin{document}
\title{\textbf{On multipartite invariant states I. \\ Unitary symmetry}} \author{Dariusz
Chru\'sci\'nski and Andrzej Kossakowski\thanks{email:
darch@phys.uni.torun.pl} }
\affiliation{Institute of Physics, Nicolaus Copernicus University,\\
Grudzi\c{a}dzka 5/7, 87--100 Toru\'n, Poland}

\begin{abstract}

We propose a natural generalization of bipartite Werner and
isotropic states to multipartite systems consisting of an
arbitrary even number of $d$-dimensional subsystems (qudits).
These generalized states are invariant under the action of local
unitary operations. We study basic properties of multipartite
invariant states: separability criteria and multi-PPT conditions.

\end{abstract}
\pacs{03.65.Ud, 03.67.-a}

\maketitle

\section{Introduction}

Symmetry plays a prominent role in modern physics. In many cases
it enables one to simplify the analysis of the corresponding
problems and very often it leads to much deeper understanding and
the most elegant mathematical formulation of the corresponding
physical theory. In Quantum Information Theory \cite{QIT} the idea
of symmetry was first applied by Werner \cite{Werner1} to
construct an important family of bipartite $d \ot d$ quantum
states which are invariant under the following local unitary
operations
\begin{equation}\label{W}
\rho \ \longrightarrow\  U\ot U \, \rho\, (U \ot U)^\dag\ ,
\end{equation}
for any $U\in U(d)$, where $U(d)$ denotes the group of unitary $d
\times d$ matrices. Another family of symmetric states (so called
isotropic states \cite{Horodecki}) is governed by the following
invariance rule
\begin{equation}\label{I}
\rho\  \longrightarrow\  U\ot \overline{U} \, \rho \, (U \ot
\overline{U})^\dag\ ,
\end{equation}
where $\overline{U}$ is the complex conjugate of $U$ in some
basis.

In the present paper we propose a natural generalization of these
two families of symmetric states to $2K$ partite quantum systems.
A generalization is straightforward: instead of 2 $d$-dimensional
systems (qudits), say Alice--Bob pair $\mathcal{H}_{AB}=
\mathcal{H}_A \ot \mathcal{H}_B$ with $\mathcal{H}_A =
\mathcal{H}_B = \Cd$, we introduce $2K$ qudits with the total
space $\mathcal{H} = \mathcal{H}_1 \ot \ldots \ot \mathcal{H}_{2K}
= (\mathbb{C}^d)^{\ot 2K}$. We may still interpret the total
system as a bipartite one with $\mathcal{H}_A =\mathcal{H}_1 \ot
\ldots \ot \mathcal{H}_{K}$ and $\mathcal{H}_B =\mathcal{H}_{K+1}
\ot \ldots \ot \mathcal{H}_{2K}$. Equivalently, we may introduce
$K$ Alices and $K$ Bobs with $\mathcal{H}_{A_i} = \mathcal{H}_i$
and $\mathcal{H}_{B_i} = \mathcal{H}_{K+i}$, respectively. Then
$\mathcal{H}_A$ and $\mathcal{H}_B$ stand for the composite $K$
Alices' and Bobs' spaces. Now, we call a  $2K$ partite quantum
state a Werner state state iff it is invariant under (\ref{W}) in
each Alice-Bob pair ${A_i} \ot {B_i}$. Similarly, the defining
property of the  generalized $2K$ partite isotropic state is that
it is invariant under (\ref{I}) in each Alice-Bob pair ${A_i} \ot
{B_i}$. Note, that for $K>1$ one has much more possibilities: the
most general invariant state is invariant under (\ref{W}) in some
pairs, say $A_1 \ot B_1, \ldots , A_L \ot B_L$ and it is invariant
under (\ref{I}) in the remaining pairs: $A_{L+1} \ot B_{L+1},
\ldots , A_K \ot B_K$. There are exactly $2^K$ different families
of invariant $2K$--partite states and for $K=1$ they reduce to the
family of Werner and isotropic states.

We analyze basic properties of these symmetric families. They are
not independent but related by a set of $2^K$ generalized partial
transpositions. Interestingly, each family gives rise to
$2^K-1$--dimensional simplex. We formulate the corresponding
multi-separability conditions and derive the generalized PPT
criterions.

A generalization of Werner states for four and three partite
system was considered in \cite{Werner2} and \cite{Werner3}. Here
we solve the problem for even number of parties in full
generality.

The symmetric states of bipartite systems proved to be very useful
in Quantum Information Theory. In particular The Peres-Horodecki
PPT criterion \cite{Peres,PPT} turns out to be the sufficient
condition for separability for symmetric states. Moreover, they
play crucial role in entanglement distillation
\cite{DIST1,DIST2,DIST3}. It is hoped that multipartite invariant
state would play similar role in multipartite composite systems.
Recently, there is a considerable effort to explore multipartite
entanglement \cite{MULTI1,MULTI2,MULTI3,MULTI4,MULTI5,MULTI6} and
symmetric states may serve as a very useful laboratory.

The paper is organized as follows: in Section \ref{2-PARTIES} we
recall basic properties of symmetric states for bipartite systems.
For pedagogical reason we first show in Section \ref{4-PARTIES}
how to generalize symmetric states for 4-partite systems and then
in Section \ref{GENERAL} we construct a general symmetric states
for an arbitrary even $2K$ number of parties.

In a forthcoming paper we present new classes of multipartite
invariant states by relaxing invariance to certain subgroups of
$U(d)$.

\section{2--partite invariant states}
\label{2-PARTIES}

\subsection{Werner state}

Werner states \cite{Werner1} play significant role in quantum
information theory. Their characteristic property is that they
commute with all unitaries of the form $U \ot U$, that is, they
are invariant under (\ref{W}):
\begin{equation}\label{}
    \mathcal{W} = U\ot U \, \mathcal{W}\, (U \ot U)^\dag\ .
\end{equation}
The space of $U\ot U$--invariant states is spanned by identity
$I^{\ot 2}$ and the flip (permutation) operator
$\mathbf{F}(\psi\ot \varphi) = \varphi \ot \psi$ defined by
\begin{equation}\label{}
    \mathbf{F} = \sum_{i,j=1}^d\, |ij\>\<ji| \ .
\end{equation}
Hence, any $U\ot U$--invariant operator may be written as $\alpha
I + \beta \mathbf{F}$. Let us introduce two projectors
\begin{equation}\label{Q}
    Q^0 = \frac 12 ( I^{\ot 2}  + \mathbf{F}) \ , \ \ \
    Q^1 = \frac 12 ( I^{\ot 2}  - \mathbf{F})\ ,
\end{equation}
i.e. $Q^0$ ($Q^1$) is the projector onto the symmetric
(anti-symmetric) subspace of $\Cd \ot \Cd$. Clearly, $Q^\alpha$
are ${U}\ot U$--invariant, $Q^\alpha Q^\beta =
\delta_{\alpha\beta}Q^\beta$, and $Q^0 + Q^1 = I^{\ot 2}$.

Now, the bipartite Werner state may written as follows
\begin{equation}\label{}
    \mathcal{W}_{\bf q} =  q_0\,
    \widetilde{Q}^0 + q_1\,\widetilde{Q}^1 \ ,
\end{equation}
where $\widetilde{Q}^\alpha = Q^\alpha/\mbox{Tr}Q^\alpha$ and the
corresponding fidelities $\mathbf{q}=(q_0,q_1)$ are given by
\begin{equation}\label{}
    q_\alpha = \mbox{Tr} ( \mathcal{W}_{\bf q}
    {Q}^\alpha) \ ,
\end{equation}
and satisfy $q_\alpha\geq 0$ together with $q_0 + q_1=1$. Werner
showed  that $\mathcal{W}_{\bf q}$ is separable iff $q_1 \leq
1/2$.

It is evident that an arbitrary bipartite state $\rho$ may be
projected  onto the $U\ot U$--invariant subspace of bipartite
Werner state by the following {\it twirl} operation:
\begin{equation}\label{cal-D}
    \mathcal{D}\rho = \int\, U\ot U \, \rho\, U^\dag \ot U^\dag\, dU\ ,
\end{equation}
where $d{U}$ is an invariant normalized Haar measure on $U(d)$,
that is, $\mathcal{D}\rho = \mathcal{W}_{\bf q}$ with fidelities
$q_\alpha = \mbox{Tr}(\rho Q^\alpha)$.

 Consider now
a partial transposition  $(\oper \ot \tau)\rho$ (we denote by
$\oper$ an identity operation acting on $M_d=$ set of $d \times d$
matrices) of a state $\rho$. Taking into account that
\begin{equation}\label{}
(\oper \ot \tau){\bf F} = d\,P^+_d\ ,
\end{equation}
where $P^+_d$ is a 1-dimensional projector corresponding to a
canonical maximally entangled state $\psi^+_d = d^{-1/2}\sum_i
|ii\>$, that is
\begin{equation}\label{}
    P^+_d = \frac 1d\, \sum_{i,j=1}^d\, |ii\>\<jj|\ ,
\end{equation}
and noting that
\begin{equation}\label{Tr-Q}
\mbox{Tr}\, Q^\alpha = \frac 12 d(d + (-1)^\alpha) \ ,
\end{equation}
one easily finds
\begin{equation}\label{Q-P}
(\oper \ot \tau)\widetilde{Q}^\alpha = \sum_{\beta=0}^1\,{\bf
X}_{\alpha\beta}\, \widetilde{P}^\beta\ ,
\end{equation}
where we introduced
\begin{equation}\label{P-alpha}
    P^1 = P^+_d\ , \ \ \ P^0 = I^{\ot 2} - P^1\ ,
\end{equation}
together with $\widetilde{P}^\alpha =
    P^\alpha/\mbox{Tr}P^\alpha$,  and the $2\times 2$ matrix $\bf X$ reads
\begin{equation}\label{X}
    {\bf X} = \frac 1d\,\left( \begin{array}{cr} d-1 & 1 \\ d+1 & -1 \end{array} \right) \
    .
\end{equation}
Note, that
\begin{equation}\label{}
    \sum_{\beta =0}^{1}\, {\bf X}_{\alpha\beta} = 1\ ,
\end{equation}
but ${\bf X}_{11} <0$ which prevents ${\bf X}$ to be a stochastic
matrix. The partial transposition of $\mathcal{W}_{\bf q}$ is
therefore given by
\begin{equation}\label{}
(\oper \ot \tau)\mathcal{W}_{\bf q} = \sum_{\alpha=0}^1\,
p'_\alpha\,\widetilde{P}^\alpha\ ,
\end{equation}
with $q'_\alpha = \sum_\beta q_\beta {\bf X}_{\beta\alpha}$.
Hence, $\mathcal{W}_{\bf q}$ is PPT iff $q'_\alpha \geq 0$ which
reproduces well known result $q_1  \leq 1/2$, i.e. Werner states
$\mathcal{W}_{\bf q}$ is separable iff it is PPT.

\subsection{Isotropic state}

Consider now another class of bipartite states -- so called
isotropic states \cite{Horodecki} -- which are invariant under
(\ref{I}), i.e.
\begin{equation}\label{U-1}
\mathcal{I} = U\ot \overline{U} \, \mathcal{I}\, (U \ot
\overline{U})^\dag\ .
\end{equation}
Note that
\begin{eqnarray}\label{U-1-U}
\lefteqn{ U\ot \overline{U} \, \rho\, (U \ot \overline{U})^\dag
}\nonumber\\ &&= (\oper \ot \tau) \Big[(U \ot U) (\oper \ot
    \tau)\rho (U \ot U)^\dag\Big]\ .
\end{eqnarray}
Let us observe that the space of $U\ot \overline{U}$--invariant
states is spanned by $P^0$ and $P^1$ defined in (\ref{P-alpha}).
Moreover, $P^\alpha
 P^\beta=\delta_{\alpha\beta}P^\beta$ and $P^0 + P^1 = I^{\ot 2}$.
Therefore, an isotropic state may be written as follows:
\begin{equation}\label{}
    \mathcal{I}_{\bf p} = \sum_{\alpha=0}^1\, p_\alpha
    \widetilde{P}^\alpha\ ,
\end{equation}
where the corresponding fidelities
\begin{equation}\label{}
    p_\alpha = \mbox{Tr} ( \mathcal{I}_{\bf p}
    {P}^\alpha) \ ,
\end{equation}
satisfy $p_\alpha\geq 0$ and $p_0 + p_1=1$.  An isotropic state is
separable iff $p_1 \leq 1/d$.

 In analogy to (\ref{cal-D}) one may define
projector into the space of $U\ot \overline{U}$--invariant states
\begin{equation}\label{cal-D-1}
    \mathcal{E}\rho = \int\, U\ot \overline{U} \, \rho\, (U \ot \overline{U})^\dag \, dU\ ,
\end{equation}
such that for any state $\rho$ one has $\mathcal{E}\rho =
\mathcal{I}_{\bf p}$ with $p_\alpha = \mbox{Tr}(\rho P^\alpha)$.
It follows from from (\ref{U-1-U}) that
\begin{equation}\label{}
    \mathcal{E} = (\oper \ot \tau) \circ \mathcal{D}\circ (\oper \ot
    \tau)\ .
\end{equation}

Finally, it is easy to show that the partial transposition $(\oper
\ot \tau)\widetilde{P}^\alpha$ is given by
\begin{equation}\label{P-Q}
(\oper \ot \tau)\widetilde{P}^\alpha = \sum_{\beta=0}^1\,{\bf
Y}_{\alpha\beta}\, \widetilde{Q}^\beta\ ,
\end{equation}
where  the $2\times 2$ matrix $\bf Y$ reads
\begin{equation}\label{Y}
    {\bf Y} = \frac 12\,\left( \begin{array}{cc} 1 & 1 \\ 1+d & 1-d \end{array} \right) \
    .
\end{equation}
Comparing  (\ref{Q-P}) and (\ref{P-Q}) it is evident that
$\mathbf{Y} = \mathbf{X}^{-1}$. Now, a state $\mathcal{I}_{\bf p}$
is PPT iff $p'_\alpha=\sum_\beta p_\beta {\bf Y}_{\beta\alpha}
\geq 0$, that is iff $p_1 \leq 1/d$. Hence, like a Werner state,
an isotropic state is separable iff it is PPT.

\section{2$\times$2--partite invariant states}
\label{4-PARTIES}

\subsection{Werner state}

Consider now the following action of the unitary group $U(d)\times
U(d)$ on 4-partite state $\rho$
\begin{equation}\label{cal-U}
   \rho\ \longrightarrow\  \mathbf{U}\ot \mathbf{U}\,\rho\,
    \mathbf{U}^\dag \ot \mathbf{U}^\dag\ ,
\end{equation}
where $\mathbf{U}=(U_1,U_2)$, with $U_i \in U(d)$ and
\[ \mathbf{U} \ot \mathbf{U} = U_1 \ot  U_2 \ot U_1 \ot
U_2 \ . \]
 The 4-dimensional space
of $\mathbf{U}\ot \mathbf{U}$--invariant states is spanned  by
\begin{equation}\label{}
    I^{\ot 4}\ , \ \ I_{1|3}^{\ot 2} \ot \mathbf{F}_{2|4} \ , \ \
    \mathbf{F}_{1|3} \ot I_{2|4}^{\ot 2}\ , \ \ \mathbf{F}_{1|3}
    \ot \mathbf{F}_{2|4} \ , \nonumber
\end{equation}
where $L_{i|j}$ denotes a bipartite operator acting on
$\mathcal{H}_i \ot \mathcal{H}_j$. Hence, for example
$I_{1|3}^{\ot 2} \ot \mathbf{F}_{2|4}$ denotes the following
operator in $\mathcal{H}_1 \ot \ldots \ot \mathcal{H}_4$:
\begin{equation}\label{}
I_{1|3}^{\ot 2} \ot \mathbf{F}_{2|4} = \sum_{i,j=1}^d\, I \ot
|i\>\<j| \ot I \ot |j\>\<i| \ . \nonumber
\end{equation}
Using Alice-Bob terminology the 4-partite operator $I_{1|3}^{\ot
2} \ot \mathbf{F}_{2|4}$ represents identity operator on the first
pair $A_1 \ot B_1$ and the operator $\mathbf{F}$ acting  on the
second pair $A_2 \ot B_2$.

 However, the more convenient
way to parameterize $\mathbf{U}\ot \mathbf{U}$--invariant subspace
is to introduce the following 4-partite orthogonal projectors:
\begin{eqnarray}\label{4Q}
\mathbf{Q}^0 &=& Q^0_{1|3} \ot Q^0_{2|4} \ , \nonumber\\
\mathbf{Q}^1 &=& Q^0_{1|3} \ot Q^1_{2|4} \ , \nonumber\\
\mathbf{Q}^2 &=& Q^1_{1|3} \ot Q^0_{2|4} \ , \\
\mathbf{Q}^3 &=& Q^1_{1|3} \ot Q^1_{2|4} \ , \nonumber
\end{eqnarray}
where $Q^\alpha$ are bipartite projectors defined in (\ref{Q}). It
is evident that $\mathbf{Q}^i$ are $U\ot U$--invariant,
$\mathbf{Q}^i \mathbf{Q}^j = \delta_{ij} \mathbf{Q}^j$, and
$\sum_{i=0}^3 \mathbf{Q}^i = I^{\ot 4}$. Now, let us introduce
more compact notation: denote by $\balpha$  a binary
$2$-dimensional vector, i.e. $\mbox{\boldmath $\alpha$}=
(\alpha_1,\alpha_2)$ with $\alpha_i \in \{0,1\}$. Clearly, any
binary vector $\mbox{\boldmath $\alpha$}$ defines an integer
number which can be written in binary notation as
$\alpha_1\alpha_2$. Using this notation the family (\ref{4Q}) may
be rewritten in a compact form as follows:
\begin{equation}\label{Q-sigma}
{\bf Q}^{ \mbox{\scriptsize\boldmath $\alpha$}} =
Q^{\alpha_1}_{1|3}
    \ot Q^{\alpha_2}_{2|4} \ .
\end{equation}
A 4-partite Werner state is defined by
\begin{equation}\label{4-Werner}
    \mathcal{W}_{\bf q}^{(2)} = \sum_{i=0}^3 q_i \widetilde{\mathbf{Q}}^i\,\equiv\, \sum_{\mbox{\scriptsize\boldmath
    $\alpha$}}\, q_{\mbox{\scriptsize\boldmath $\alpha$}}
    \widetilde{{\mathbf{Q}}}^{\mbox{\scriptsize\boldmath $\alpha$}}\ ,
\end{equation}
where $\widetilde{\mathbf{Q}}^{\mbox{\scriptsize\boldmath
$\alpha$}} = {\mathbf{Q}}^{\mbox{\scriptsize\boldmath
$\alpha$}}/\mbox{Tr}{\mathbf{Q}}^{\mbox{\scriptsize\boldmath
$\alpha$}}$, and the corresponding fidelities
\begin{equation}\label{}
    q_{\mbox{\scriptsize\boldmath $\alpha$}} =
    \mbox{Tr}(\mathcal{W}_{\bf q}^{(2)} {\mathbf{Q}}^{\mbox{\scriptsize\boldmath
    $\alpha$}}) \geq 0 \ ,
\end{equation}
satisfy $ \sum_{\mbox{\scriptsize\boldmath $\alpha$}}
q_{\mbox{\scriptsize\boldmath $\alpha$}} = 1$. Note, that
\begin{equation}\label{}
\widetilde{\mathbf{Q}}^{\mbox{\scriptsize\boldmath $\alpha$}} =
\widetilde{Q}^{\alpha_1}_{1|3}
    \ot \widetilde{Q}^{\alpha_2}_{2|4} \ ,
\end{equation}
and hence, using (\ref{Tr-Q}), one obtains
\begin{eqnarray}\label{}
    \mbox{Tr}\,{\mathbf{Q}}^\sbalpha &=& \left(\frac{d}{2}\right)^2 (d + (-1)^{\alpha_1})(d + (-1)^{\alpha_2})
    \nonumber \\ &=& \left(\frac{d}{2}\right)^2 (d -1)^{|\sbalpha|}(d +
    1)^{2-|\sbalpha|}\ ,
\end{eqnarray}
where $|\!\!\balpha|= \alpha_1 + \alpha_2 \in \{ 0,1,2\}$.

This way the space of  4-partite-Werner states  defines
3--dimensional simplex. The vertices of this simplex correspond to
$\widetilde{\mathbf{Q}}^{ \mbox{\scriptsize\boldmath $\alpha$}}$.

It is evident that an arbitrary 4-partite state $\rho$ may be
projected  onto the $\mathbf{U}\ot \mathbf{U}$--invariant subspace
of 4-partite Werner state by the following {\it twirl} operation:
\begin{equation}\label{cal-D-2}
    \mathcal{D}^{(2)}\rho = \int\, \mathbf{U}\ot \mathbf{U}\,\rho\,
    \mathbf{U}^\dag \ot \mathbf{U}^\dag\, d\mathbf{U}\ ,
\end{equation}
where $d{\bf U}=dU_1dU_2$ is an invariant normalized Haar measure
on $U(d)^2$, that is, $\mathcal{D}^{(2)}\rho = \mathcal{W}_{\bf
q}^{(2)}$ with fidelities $q_\sbalpha = \mbox{Tr}(\rho {\bf
Q}^{\sbalpha})$.

To find the corresponding separability criteria note that
$\mathcal{W}_{\bf q}^{(2)}$ is separable iff there exists a
separable state $\rho$ such that $\mathcal{D}^{(2)}\rho =
\mathcal{W}_{\bf q}^{(2)}$. Let $\rho$ be an extremal  separable
state of the form
\begin{equation}\label{PPPP}
    \rho = P_{\psi_1} \ot P_{\psi_2} \ot P_{\varphi_1} \ot P_{\varphi_2}
    \ ,
\end{equation}
where $P_{\psi} = |\psi\>\<\psi|$, and $\psi_i,\varphi_i$ are
normalized vectors in $\Cd$. An arbitrary 4-separable state is a
convex combination of the extremal states of the form
({\ref{PPPP}). One easily finds for fidelities $ \mbox{Tr}(\rho
{\bf Q}^{\sbalpha})$:
\begin{eqnarray}\label{4qa}
q_0 &=& q_{(00)} = \frac 14(1+a_1)(1+a_2) \ , \nonumber \\
q_1 &=& q_{(01)} =  \frac 14 (1+a_1)(1-a_2) \ , \nonumber \\
q_2 &=& q_{(10)} =  \frac 14 (1-a_1)(1+a_2) \ , \\
 q_3 &=&
q_{(11)} = \frac 14(1-a_1)(1-a_2) \ , \nonumber
\end{eqnarray}
with
\begin{equation}\label{a-12}
    a_1 = |\<\psi_1|\varphi_1\>|^2\ , \ \ \ \ \  a_2 = |\<\psi_2|\varphi_2\>|^2\
    .
\end{equation}
These formulae may be rewritten in a compact form as follows:
\begin{equation}\label{}
    q_{\sbalpha} = \frac 14 ( 1 +
    (-1)^{\alpha_1}a_1)(1+(-1)^{\alpha_2}a_2) \ .
\end{equation}
Now, since $a_i \leq 1$, the projection $\mathcal{D}^{(2)}$ of the
convex hull of extremal separable states gives therefore
\begin{equation}\label{WS-1}
q_{00} \leq 1\ , \ \ \ \     q_{01}, q_{10} \leq \frac 12\ , \ \ \
\ q_{11} \leq \frac 14\ ,
\end{equation}
together with
\begin{equation}\label{q<q<q}
    q_{11} \leq q_{01}, q_{10} \leq q_{00}\ .
\end{equation}
Note, that using binary notation equations (\ref{WS-1}) may be
compactly rewritten as follows
\begin{equation}\label{4qa||}
    q_{\sbalpha} \leq \frac{1}{2^{|\sbalpha|}}\ .
\end{equation}

\subsection{Isotropic state}

Now, in analogy to the bipartite case we may define a 4-partite
isotropic state $\mathcal{I}_{\bf p}^{(2)}$ which is invariant
under
\begin{equation}\label{cal-U*}
    \rho' = \mathbf{U}\ot \overline{\mathbf{U}}\,\rho\,
   ( \mathbf{U} \ot \overline{\mathbf{U}})^\dag\ ,
\end{equation}
with $\mathbf{U} \ot \overline{\mathbf{U}} = U_1 \ot  U_2 \ot
\overline{U}_1 \ot \overline{U}_2$. The recipe is very simple:
starting from (\ref{4Q}) we may replace both $Q$'s by ${P}$'s
defined in (\ref{P-alpha}). One obtains the following family of
orthogonal projectors:
\begin{eqnarray}\label{4P}
\mathbf{P}^0 &=& P^0_{1|3} \ot P^0_{2|4} \ , \nonumber\\
\mathbf{P}^1 &=& P^0_{1|3} \ot P^1_{2|4} \ , \nonumber\\
\mathbf{P}^2 &=& P^1_{1|3} \ot P^0_{2|4} \ , \\
\mathbf{P}^3 &=& P^1_{1|3} \ot P^1_{2|4} \ . \nonumber
\end{eqnarray}
It is evident that
\begin{equation}\label{}
\mathbf{U}\ot \overline{\mathbf{U}}\,\mathbf{P}^i\,
   ( \mathbf{U} \ot \overline{\mathbf{U}})^\dag = \mathbf{P}^i\ .
\end{equation}
Moreover, one has $\mathbf{P}^i \mathbf{P}^j = \delta_{ij}
\mathbf{P}^j$, and $\sum_{i=0}^3 \mathbf{P}^i = I^{\ot 4}$.
Therefore, any $ \mathbf{U}\ot \overline{\mathbf{U}}$--invariant
state may be written as follows
\begin{equation}\label{4-Isotropic}
    \mathcal{I}_{\bf p}^{(2)} = \sum_{i=0}^3 p_i \widetilde{\mathbf{P}}^i\,\equiv\, \sum_{\mbox{\scriptsize\boldmath
    $\alpha$}}\, p_{\mbox{\scriptsize\boldmath $\alpha$}}
    \widetilde{{\mathbf{P}}}^{\mbox{\scriptsize\boldmath $\alpha$}}\ ,
\end{equation}
where as usual $\widetilde{A}= A/\mbox{Tr}A$, and
\begin{equation}\label{P-sigma}
{\bf P}^{ \mbox{\scriptsize\boldmath $\alpha$}} =
P^{\alpha_1}_{1|3}
    \ot P^{\alpha_2}_{2|4} \ .
\end{equation}
One easily finds
\begin{eqnarray}\label{}
    \mbox{Tr}\, {\mathbf{P}}^\sbalpha = (d^2-1)^{2-|\sbalpha|}\ .
\end{eqnarray}
 The fidelities
\begin{equation}\label{}
    p_{\mbox{\scriptsize\boldmath $\alpha$}} =
    \mbox{Tr}(\mathcal{I}_{\bf p}^{(2)} {\mathbf{P}}^{\mbox{\scriptsize\boldmath
    $\alpha$}}) \geq 0 \ ,
\end{equation}
satisfy $ \sum_{\mbox{\scriptsize\boldmath $\alpha$}}
p_{\mbox{\scriptsize\boldmath $\alpha$}} = 1$.

Denote by $\mathcal{E}^{(2)}$ on orthogonal projector onto the
space of $ \mathbf{U}\ot \overline{\mathbf{U}}$--invariant states
\begin{equation}\label{cal-E-2}
    \mathcal{E}^{(2)}\rho = \int\, \mathbf{U}\ot \overline{\mathbf{U}}\,\rho\,
    \mathbf{U}^\dag \ot  \overline{\mathbf{U}}^\dag\, d\mathbf{U}\
    .
\end{equation}
It is evident that
\begin{eqnarray}\label{}
\mathcal{E}^{(2)} =  (\oper \ot \oper \ot \tau \ot \tau)  \circ
\mathcal{D}^{(2)} \circ  (\oper \ot \oper \ot \tau \ot \tau)\ .
\end{eqnarray}
Now, an isotropic state $\mathcal{I}_{\bf p}^{(2)}$ is separable
iff there exists a separable state $\rho$ such that
$\mathcal{E}^{(2)}\rho = \mathcal{I}_{\bf p}^{(2)}$. Let us
consider  an extremal  separable state $(\oper \ot \oper \ot \tau
\ot \tau)\rho$ with $\rho$ defined in (\ref{PPPP}), i.e. i.e.
\begin{equation}\label{}
  (\oper \ot \oper \ot \tau \ot \tau)\rho = P_{\psi_1} \ot P_{\psi_2} \ot P^T_{\varphi_1} \ot P^T_{\varphi_2}
    \ ,
\end{equation}
and define the isotropic state $\mathcal{E}^{(2)}( P_{\psi_1} \ot
P_{\psi_2} \ot P^T_{\varphi_1} \ot P^T_{\varphi_2})$. One easily
finds for fidelities:
\begin{eqnarray}\label{4pb}
p_0 &=& p_{(00)} = (1-b_1)(1-b_2) \ , \nonumber \\
p_1 &=& p_{(01)} = b_1(1-b_2) \ , \nonumber \\
p_2 &=& p_{(10)} = (1-b_1)b_2 \ , \\
 p_3 &=&
p_{(11)} =  b_1b_2 \ , \nonumber
\end{eqnarray}
or equivalently
\begin{equation}\label{}
    p_{\sbalpha} = (1-[\alpha_1 + (-1)^{\alpha_1}b_1]) (1-[\alpha_2 +
    (-1)^{\alpha_2}b_2])\ ,
\end{equation}
 with
\begin{equation}\label{}
    b_i = \frac{a_i}{d} = \frac{|\<\psi_i|\varphi_i\>|^2}{d}\ .
\end{equation}
Now, since $b_i \leq 1/d$, the projection $\mathcal{E}^{(2)}$ of
the convex hull of extremal separable states gives therefore
\begin{equation}\label{IS-1}
p_{00} \leq 1\ , \ \ \ \     p_{01}, p_{10} \leq \frac 1d\ , \ \ \
\ p_{11} \leq \frac{1}{d^2}\ ,
\end{equation}
or more compactly in binary notation
\begin{equation}\label{4pb||}
    p_{\sbalpha} \leq \frac{1}{d^{|\sbalpha|}}\ ,
\end{equation}
and
\begin{equation}\label{IS-2}
    p_{11} \leq p_{01}, p_{10} \leq p_{00}\ .
\end{equation}

\subsection{$\bsigma$--invariant states}

Let us observe that in $\mathcal{H}_A \ot \mathcal{H}_B$ we may
define not only the partial transposition $\oper \ot \oper \ot
\tau \ot \tau$ considered in the previous Section but also the
following ones:
\begin{eqnarray}\label{}
\tau_1 &=& (\oper \ot \oper \ot \oper \ot \tau) \ ,   \\
\tau_2 &=& (\oper \ot \oper \ot \tau \ot \oper) \ .
\end{eqnarray}
All partial transpositions in Alice-Bob system may be conveniently
denoted by
\begin{equation}\label{}
    \tau_{\sbsigma} = \oper \ot \oper \ot \tau^{\sigma_1} \ot
    \tau^{\sigma_2} \ ,
\end{equation}
where
\begin{equation}\label{tau-alpha}
\tau^{\alpha} = \left\{ \begin{array}{ll} \oper  \ , & \ \
\alpha=0 \\  \tau \ ,& \ \ \alpha=1
\end{array} \right. \ .
\end{equation}
Clearly, for $\bsigma =(0,0)$ one has trivial operation
$\tau_{(00)} = \oper^{\ot 4}$, whereas $ \tau_{(01)} = \tau_1$,
$\tau_{(10)} = \tau_2$ and $\tau_{(11)}$ reproduces double partial
transposition $\oper \ot \oper \ot \tau \ot \tau$.

We call a 4-partite state $\rho$ a $\bsigma$--invariant iff
$\tau_{\sbsigma}\rho$ is $\mathbf{U} \ot \mathbf{U}$--invariant
i.e.
\begin{equation}\label{}
 (\mathbf{U} \ot \mathbf{U}) (\tau_{\sbsigma}\rho) (\mathbf{U}
\ot \mathbf{U})^\dag = \tau_{\sbsigma}\rho \ .
\end{equation}
To characterize $\bsigma$--invariant states
  let us define the following families of
projectors:
\begin{eqnarray}\label{4Pi1}
\mathbf{\Pi}_{(1)}^0 &=& Q^0_{1|3} \ot P^0_{2|4} \ , \nonumber\\
\mathbf{\Pi}_{(1)}^1 &=& Q^0_{1|3} \ot P^1_{2|4} \ , \nonumber\\
\mathbf{\Pi}_{(1)}^2 &=& Q^1_{1|3} \ot P^0_{2|4} \ , \\
\mathbf{\Pi}_{(1)}^3 &=& Q^1_{1|3} \ot P^1_{2|4} \ , \nonumber
\end{eqnarray}
and
\begin{eqnarray}\label{4Qi1}
\mathbf{\Pi}_{(2)}^0 &=& P^0_{1|3} \ot Q^0_{2|4} \ , \nonumber\\
\mathbf{\Pi}_{(2)}^1 &=& P^0_{1|3} \ot Q^1_{2|4} \ , \nonumber\\
\mathbf{\Pi}_{(2)}^2 &=& P^1_{1|3} \ot Q^0_{2|4} \ , \\
\mathbf{\Pi}_{(2)}^3 &=& P^1_{1|3} \ot Q^1_{2|4} \ . \nonumber
\end{eqnarray}
Let us observe that  4 families: $\mathbf{Q}^\sbalpha$,
$\mathbf{P}^\sbalpha$,  $\mathbf{\Pi}_{(1)}^\sbalpha$ and
$\mathbf{\Pi}_{(2)}^\sbalpha$ may be compactly written as
\begin{equation}\label{Pi-4}
\mathbf{\Pi}_{(\sbsigma)}^\sbalpha =
\Pi^{\alpha_1}_{(\sigma_1)1|3} \ot \Pi^{\alpha_2}_{(\sigma_2)2|4}
\ ,
\end{equation}
where
\begin{equation}\label{Pi-QP}
\Pi^{\alpha}_{(\sigma)} = \left\{ \begin{array}{ll} {Q}^{\alpha} \
, & \ \ \sigma=0
\\ P^{\alpha} \ ,& \ \ \sigma=1
\end{array} \right. \ ,
\end{equation}
that is,
\begin{eqnarray}\label{}
    \mathbf{\Pi}_{(00)}^\sbalpha &=& \mathbf{Q}^\sbalpha\ , \ \ \
    \ \ \
 \mathbf{\Pi}_{(01)}^\sbalpha\, =\, \mathbf{\Pi}_{(1)}^\sbalpha\
 ,\nonumber \\
 \mathbf{\Pi}_{(10)}^\sbalpha &=& \mathbf{\Pi}_{(2)}^\sbalpha\ ,\
 \ \ \ \,
  \mathbf{\Pi}_{(11)}^\sbalpha\, =\, \mathbf{P}^\sbalpha\ .\nonumber
\end{eqnarray}
One easily shows that
\begin{enumerate}
\item $\ \ \mathbf{\Pi}_{(\sbsigma)}^\sbalpha$ are
$\bsigma$--invariant,

\item $\ \ \mathbf{\Pi}_{(\sbsigma)}^\sbalpha\cdot
\mathbf{\Pi}_{(\sbsigma)}^\sbbeta = \delta_{\sbalpha \sbbeta}\,
\mathbf{\Pi}_{(\sbsigma)}^\sbbeta$,

\item $\ \ \sum_{\sbalpha}\, \mathbf{\Pi}_{(\sbsigma)}^\sbalpha\, = \,
\oper^{\ot 4}\ . $

\end{enumerate}

It is therefore clear that any $\bsigma$--invariant state may be
written as follows:
\begin{equation}\label{}
    \mathcal{I}^{(\sbsigma)}_{\bf f} = \sum_{\sbalpha}\,
    f^{(\sbsigma)}_{\sbalpha}
    \widetilde{\mathbf{\Pi}}^{\sbalpha}_{(\sbsigma)}\ ,
\end{equation}
where the corresponding fidelities
\begin{equation}\label{}
f^{(\sbsigma)}_{\sbalpha} =
\mbox{Tr}(\mathcal{I}^{(\sbsigma)}_{\bf
f}{\mathbf{\Pi}}^{\sbalpha}_{(\sbsigma)}) \ ,
\end{equation}
satisfy $ \sum_{\sbalpha}\, f^{(\sbsigma)}_{\sbalpha}  = 1$.
Clearly, one has $f^{(00)}_{\sbalpha} = q_\sbalpha$ and
$f^{(11)}_{\sbalpha} = p_\sbalpha$.

Now, to check for separability conditions note that $
\mathcal{I}^{(\sbsigma)}_{\bf f}$ is separable iff there exists a
separable state $\rho$ such that $\mathcal{D}^{(2)}_\sbsigma \rho
$ is separable, where
\begin{equation}\label{}
    \mathcal{D}^{(2)}_\sbsigma = \tau_\sbsigma \circ
    \mathcal{D}^{(2)} \circ \tau_\sbsigma\ ,
\end{equation}
denotes the projector onto the subspace of $\bsigma$--invariant
states. It is evident that $\mathcal{D}^{(2)}_{(00)} =
\mathcal{D}^{(2)}$ and $\mathcal{D}^{(2)}_{(11)} =
\mathcal{E}^{(2)}$. In analogy to (\ref{4qa}) and (\ref{4pb}) one
easily finds for fidelities corresponding to $
\mathcal{D}^{(2)}_{(01)}(\rho)$ with $\rho$ being en extremal
separable state (\ref{PPPP}):
\begin{eqnarray}\label{4ab}
 f^{(01)}_{(00)} &=& \frac 12(1+a_1)(1-b_2) \ , \nonumber \\
f^{(01)}_{(01)} &=&  \frac 12 (1+a_1)b_2 \ , \nonumber \\
f^{(01)}_{(10)} &=&  \frac 12 (1-a_1)(1-b_2) \ , \\
f^{(01)}_{(11)} &=& \frac 12(1-a_1)b_2 \ , \nonumber
\end{eqnarray}
and similarly for $ \mathcal{D}^{(2)}_{(10)}(\rho)$
\begin{eqnarray}\label{4ba}
 f^{(10)}_{(00)} &=& \frac 12\,(1-b_1)(1+a_2) \ , \nonumber \\
f^{(10)}_{(01)} &=&  \frac 12\, (1-b_1)(1-a_2) \ , \nonumber \\
f^{(10)}_{(10)} &=&  \frac 12\, b_1(1+a_2) \ , \\
f^{(10)}_{(11)} &=& \frac 12\, b_1(1+a_2) \ . \nonumber
\end{eqnarray}
The projection $\mathcal{D}^{(2)}_\sbsigma$ of the convex hull of
extremal separable states gives therefore
\begin{equation}\label{4f||}
    f^{(\sbsigma)}_{\sbalpha } \leq \frac{1}{2^{|\sbalpha|}}
    \left(\frac{2}{d}\right)^{\!|\sbsigma\sbalpha|}\ ,
\end{equation}
where $\bsigma\!\!\balpha=(\sigma_1\alpha_1,\sigma_2\alpha_2)$,
and
\begin{equation}\label{}
f^{(\sbsigma)}_{\sbalpha } \leq  f^{(\sbsigma)}_{\sbbeta }\ , \ \
\ \mbox{for} \ \ \ |\!\!\balpha|> |\!\!\bbeta|\ ,
\end{equation}
which generalize (\ref{q<q<q})--(\ref{4qa||}) and
(\ref{4pb||})--(\ref{IS-2}).

\subsection{$\bsigma$--PPT states}

We call a 4-partite state $\rho$ in $\mathcal{H}_A \ot
\mathcal{H}_B = \mathcal{H}_{A_1} \ot \mathcal{H}_{A_2} \ot
\mathcal{H}_{B_1} \ot \mathcal{H}_{B_2}$ a $\bsigma$--PPT iff
\begin{equation}\label{}
    \tau_{\sbsigma}\rho \geq 0\ .
\end{equation}
Now, if $O$ is $\bnu$--invariant operator in $\mathcal{H}_A \ot
\mathcal{H}_B$, then $\tau_\sbmu O$ is $(\!\bmu \oplus\!\!
\bnu)$--invariant, where $\oplus$ denotes addition mod 2. Writing
$O$ as
\begin{equation}\label{}
O = \sum_\sbalpha\, o_\sbalpha
\widetilde{\mathbf{\Pi}}^\sbalpha_{(\sbnu)}\ ,
\end{equation}
one has
\begin{equation}\label{}
\tau_\sbmu O = \sum_\sbalpha\, o_\sbalpha \tau_\sbmu
\widetilde{\mathbf{\Pi}}^\sbalpha_{(\sbnu)}\ .
\end{equation}
One easily computes the $\bmu$--partial transposition of
$\widetilde{\mathbf{\Pi}}^\sbalpha_{(\sbnu)}$:
\begin{equation}\label{}
\tau_\sbmu\widetilde{\mathbf{\Pi}}^\sbalpha_{(\sbnu)} =
\sum_\sbbeta\, \mathbf{Z}^{\sbalpha\sbbeta}_{(\sbmu|\sbnu)}\,
\widetilde{\mathbf{\Pi}}^\sbbeta_{(\sbmu \oplus \sbnu)}\ ,
\end{equation}
where the $4\times 4$ matrix $ \mathbf{Z}_{(\sbmu|\sbnu)}$ is
defined as follows:
\begin{equation}\label{}
\mathbf{Z}_{(\sbmu|\sbnu)} = \mathbf{Z}_{(\mu_1|\nu_1)} \ot
\mathbf{Z}_{(\mu_2|\nu_2)}\ ,
\end{equation}
with
\begin{equation}\label{}
\mathbf{Z}_{(\mu|\nu)} = \left\{ \begin{array}{ll} \mathbf{I} \ ,
& \ \ \mu=0\ , \ \ \nu =0,1
\\  \mathbf{X} \ ,& \ \ \mu=1\ , \ \ \nu=0\\
\mathbf{Y} \ ,& \ \ \mu=1\ , \ \ \nu=1
\end{array} \right. \ ,
\end{equation}
and $\mathbf{I}$ denotes $2\times 2$ unit matrix. Matrices
$\mathbf{X}$ and $\mathbf{Y}$ are defined in (\ref{X}) and
(\ref{Y}), respectively. The corresponding matrix elements are
defined in an obvious way
\[  (\mathbf{A}\ot \mathbf{B})^{\sbalpha\sbbeta} =
\mathbf{A}^{\alpha_1\beta_1} \, \mathbf{B}^{\alpha_2\beta_2}\ .
\]
The structure of $\mathbf{Z}_{(\mu|\nu)}$ is encoded into the
following table:
\[ \begin{array}{|c||c|c|c|c|}
\hline \bmu\backslash\bnu & (00) & (01) & (10) & (11) \\
\hline \hline (00) & \mathbf{I} \ot \mathbf{I} & \mathbf{I} \ot
\mathbf{I} & \mathbf{I} \ot \mathbf{I} & \mathbf{I} \ot \mathbf{I}
 \\
\hline (01) & \mathbf{I} \ot \mathbf{X} & \mathbf{I} \ot
\mathbf{Y} & \mathbf{I} \ot \mathbf{X} & \mathbf{I} \ot \mathbf{Y}
 \\
 \hline (10) & \mathbf{X} \ot \mathbf{I} & \mathbf{X} \ot
\mathbf{I} & \mathbf{Y} \ot \mathbf{I} & \mathbf{Y} \ot \mathbf{I}
 \\
  \hline (11) & \mathbf{X} \ot \mathbf{X} & \mathbf{X} \ot
\mathbf{Y} & \mathbf{Y} \ot \mathbf{X} & \mathbf{Y} \ot \mathbf{Y}
 \\
\hline
\end{array} \]

Now, if $\bnu$--invariant operator $O$ is semi-positive, i.e.
$o_\sbalpha \geq 0$, then $O$ is $\bmu$--PPT iff
\begin{equation}\label{}
    \sum_\sbbeta\, o_\sbbeta\,
    \mathbf{Z}^{\sbbeta\sbalpha}_{(\sbmu|\sbnu)} \geq 0 \ ,
\end{equation}
for all binary 2-vectors $\!\balpha$.

In particular one may look for the $\bsigma$--PPT conditions for
the 4-partite Werner state. One easily finds that

\begin{enumerate}

\item  $\mathcal{W}_{\bf q}$ is $(01)$--PPT iff
\begin{equation}\label{01PPT}
    q_{00} \geq q_{01} \ , \ \ \ q_{10} \geq q_{11} \ ,
\end{equation}

\item  $\mathcal{W}_{\bf q}$ is $(10)$--PPT iff
\begin{equation}\label{10PPT}
    q_{00} \geq q_{10} \ , \ \ \ q_{01} \geq q_{11} \ ,
\end{equation}

\item  $\mathcal{W}_{\bf q}$ is $(11)$--PPT iff
\begin{eqnarray}\label{11PPT}
(d-1)(q_{00} - q_{01}) + (d+1)(q_{10}  - q_{11}) &\geq& 0 \ ,
\nonumber \\
(d-1)(q_{00} - q_{10}) + (d+1)(q_{01}  - q_{11}) &\geq& 0 \ , \\
(q_{00} + q_{11}) - (q_{01} + q_{10}) &\geq& 0 \ . \nonumber
\end{eqnarray}
\end{enumerate}
 Note that PPT
conditions (\ref{01PPT})--(\ref{11PPT}) imply
\begin{equation}\label{qqqq}
    q_{11} \leq q_{01}, q_{10} \leq q_{00} \ ,
\end{equation}
which reproduces (\ref{q<q<q}), together with
\begin{equation}\label{}
q_{01} + q_{10} \leq q_{00} + q_{11}\ ,
\end{equation}
which is equivalent to
\begin{equation}\label{01-10-12}
q_{01} + q_{10} \leq \frac 12\ .
\end{equation}
Now, (\ref{qqqq}) and (\ref{01-10-12}) imply
\begin{equation}\label{}
    2q_{11} \leq q_{01} + q_{10} \leq \frac 12\ ,
\end{equation}
and hence
\begin{equation}\label{}
    q_{11} \leq \frac 14\ ,
\end{equation}
which, together with
\begin{equation}\label{}
q_{01} , q_{10} \leq \frac 12\ ,
\end{equation}
reproduces (\ref{4qa||}). This shows that 4-partite Werner state
is 4-separable iff it is $\bsigma$--PPT for all binary vectors
$\bsigma$.
 Interestingly, one may prove (see Appendix) that 4-partite
Werner state is $12|34$ (or $A\ot B$) bi-separable iff it is
$(11)$--PPT.

One may perform similar analysis for other invariant states.
Again, a $\bmu$-invariant state is 4-separable iff it is
$\bnu$--PPT for all binary vectors $\bnu$. It is $A \ot B$
bi-separable iff it is $(11)$--PPT.

\subsection{Reductions}

It is clear that reducing 4-partite invariant state with respect
to the  pair $A_1 \ot B_1$ ($A_2 \ot B_2$) one obtains bipartite
invariant state of $A_2 \ot B_2$ ($A_1 \ot B_1$). One easily finds
\begin{equation}\label{}
\mbox{Tr}_{13}\,  \mathcal{W}^{(2)}_{\bf q} = \mathcal{W}_{\bf
q'}\ ,
\end{equation}
with
\begin{equation}\label{}
    q'_\alpha = \sum_\beta\, q_{(\beta\alpha)}\ .
\end{equation}
Similarly,
\begin{equation}\label{}
\mbox{Tr}_{24}\,  \mathcal{W}^{(2)}_{\bf q} = \mathcal{W}_{\bf
q''}\ ,
\end{equation}
with
\begin{equation}\label{}
    q''_\alpha = \sum_\beta\, q_{(\alpha\beta)}\ .
\end{equation}
This observation may be easily generalized to an arbitrary
4-partite invariant state $\mathcal{I}^{(\sbsigma)}_{\bf f}$:
\begin{equation}\label{}
    \mbox{Tr}_{13}\, \mathcal{I}^{(\sbsigma)}_{\bf f} =
    \sum_{\alpha_2}\, f_{\alpha_2}\, \Pi^{\alpha_2}_{(\sigma_2)}\
    ,
\end{equation}
where $\Pi^{\alpha}_{(\sigma)}$ is defined in (\ref{Pi-QP}) and
\begin{equation}\label{}
f_{\alpha_2} = \sum_{\alpha_1} f_{(\alpha_1,\alpha_2)}\ .
\end{equation}
Finally, let us observe that a reduction with respect to any other
pair produces maximally mixed state of the remaining pair, e.g.
\begin{equation}\label{}
\mbox{Tr}_{12}\, \mathcal{I}^{(\sbsigma)}_{\bf f} = I^{\ot
2}_{3|4}\ .
\end{equation}

\section{$2K$--partite invariant states}
\label{GENERAL}

\subsection{General $\bsigma$--invariant state}

Consider now $2K$--partite system and define the following action
of $K$ copies of U$(d)$:
\begin{equation}\label{cal-U**}
 \rho' = \mathbf{U}\ot \mathbf{U}\,\rho\,
    \mathbf{U}^\dag \ot \mathbf{U}^\dag\ ,
\end{equation}
where $\mathbf{U}=(U_1,\ldots,U_K)$ with $U_i \in U(d)$ and
\[ \mathbf{U} \ot \mathbf{U} = U_1 \ot \ldots U_K \ot U_1 \ot \ldots
U_K \ .  \] A state $\rho$ is $\mathbf{U}\ot
\mathbf{U}$--invariant iff
\[ \mathbf{U}\ot \mathbf{U}\,\rho\, =\, \rho\, \mathbf{U}\ot \mathbf{U}\ ,
\]
for any $\mathbf{U}\in U(d)^K$. Denote by ${\cal D}^{(K)}$ the
corresponding projector onto the space of $\mathbf{U}\ot
\mathbf{U}$--invariant states
\begin{equation}\label{}
    \mathcal{D}^{(K)}\rho = \int\, d\mathbf{U}\,\mathbf{U}\ot \mathbf{U}\,\rho\,
    \mathbf{U}^\dag \ot \mathbf{U}^\dag\ ,
\end{equation}
with $d\mathbf{U}= dU_1\ldots dU_K$ being an  normalized invariant
Haar measure on $U(d)^K$.

 Now, let $\mbox{\boldmath $\sigma$}$ be a binary $K$-dimensional vector, i.e.
$\mbox{\boldmath $\sigma$}= (\sigma_1,\ldots,\sigma_K)$ with
$\sigma_j \in \{0,1\}$. For any $\bsigma$ one may define
$\bsigma$--partial transposition on $\mathcal{H}_A \ot
\mathcal{H}_B$ as follows:
\begin{equation}\label{}
    \tau_\sbsigma = \oper^{\ot K} \ot \tau^{\sigma_1} \ot \ldots
    \ot \tau^{\sigma_K} \ ,
\end{equation}
where $\tau^\alpha$ is defined in (\ref{tau-alpha}). We call a
state $\rho$ $\bsigma$--invariant iff $\tau_\sbsigma\rho$ is
$\mathbf{U}\ot \mathbf{U}$--invariant. The corresponding projector
$\mathcal{D}^{(K)}_\sbsigma$ onto the space of
$\bsigma$--invariant states reads
\begin{equation}\label{}
    \mathcal{D}^{(K)}_\sbsigma = \tau_\sbsigma \circ
    \mathcal{D}^{(K)} \circ \tau_\sbsigma\ .
\end{equation}
To parameterize the space of $\bsigma$--invariant states let us
introduce the following family of projectors:
\begin{equation}\label{Pi-2K}
\mathbf{\Pi}_{(\sbsigma)}^\sbalpha =
\Pi^{\alpha_1}_{(\sigma_1)1|K+1} \ot \ldots \ot
\Pi^{\alpha_K}_{(\sigma_K)K|2K} \ ,
\end{equation}
where $\Pi^{\alpha_i}_{(\sigma_i)}$ are defined in (\ref{Pi-QP}).
It generalizes 4-partite family (\ref{Pi-4}). Note that we have
$2^K$ families parameterized by $\bsigma$ each containing $2^K$
elements.

One easily shows that
\begin{enumerate}
\item $\ \ \mathbf{\Pi}_{(\sbsigma)}^\sbalpha$ are
$\bsigma$--invariant,

\item $\ \ \mathbf{\Pi}_{(\sbsigma)}^\sbalpha\cdot
\mathbf{\Pi}_{(\sbsigma)}^\sbbeta = \delta_{\sbalpha \sbbeta}\,
\mathbf{\Pi}_{(\sbsigma)}^\sbbeta$,

\item $\ \ \sum_{\sbalpha}\, \mathbf{\Pi}_{(\sbsigma)}^\sbalpha\, = \,
\oper^{\ot 2K}\ . $

\end{enumerate}

It is therefore clear that any $\bsigma$--invariant state may be
written as follows:
\begin{equation}\label{}
    \mathcal{I}^{(\sbsigma)}_{\bf f} = \sum_{\sbalpha}\,
    f^{(\sbsigma)}_{\sbalpha}
    \widetilde{\mathbf{\Pi}}^{\sbalpha}_{(\sbsigma)}\ ,
\end{equation}
where the corresponding fidelities
\begin{equation}\label{}
f^{(\sbsigma)}_{\sbalpha} =
\mbox{Tr}(\mathcal{I}^{(\sbsigma)}_{\bf
f}{\mathbf{\Pi}}^{\sbalpha}_{(\sbsigma)}) \ ,
\end{equation}
satisfy $ \sum_{\sbalpha}\, f^{(\sbsigma)}_{\sbalpha}  = 1$.
Hence, the space of $\bsigma$--invariant states gives rise to a
$(2^K-1)$--dimensional simplex.

In particular for $\bsigma=(0,\ldots,0)$ one obtains a
$2K$-partite Werner state
\begin{equation}\label{}
    \mathcal{W}_{\bf q}^{(K)} = \sum_{\sbalpha}\, q_\sbalpha
    \widetilde{\mathbf{Q}}^\sbalpha\ ,
\end{equation}
with
\begin{equation}\label{}
    \widetilde{\mathbf{Q}}^\sbalpha =
\widetilde{Q}^{\alpha_1}_{1|K+1} \ot \ldots \ot
\widetilde{Q}^{\alpha_K}_{K|2K} \ .
\end{equation}
On the other hand for $\bsigma=(1,\ldots,1)$ one obtains
$\mathbf{U}\ot \overline{\mathbf{U}}$--invariant  $2K$-partite
isotropic state
\begin{equation}\label{}
    \mathcal{I}_{\bf p}^{(K)} = \sum_{\sbalpha}\, p_\sbalpha
    \widetilde{\mathbf{P}}^\sbalpha\ ,
\end{equation}
with
\begin{equation}\label{}
    \widetilde{\mathbf{P}}^\sbalpha =
\widetilde{P}^{\alpha_1}_{1|K+1} \ot \ldots \ot
\widetilde{P}^{\alpha_K}_{K|2K} \ .
\end{equation}

\subsection{Separability}

To find the corresponding separability conditions for
$\bsigma$--invariant states let us consider a multi-separable
state
\begin{equation}\label{}
    \rho_\sbsigma = \tau_\sbsigma\rho\ ,
\end{equation}
with $\rho$ being a product state
\begin{equation}\label{}
    \rho = P_{\psi_1} \ot \ldots \ot P_{\psi_K} \ot  P_{\varphi_1} \ot \ldots \ot
    P_{\varphi_K}\ .
\end{equation}
One easily computes the corresponding fidelities
\begin{equation}\label{}
    f^{(\sbsigma)}_\sbalpha = \mbox{Tr}( \rho_\sbsigma
    \mathbf{\Pi}^\sbalpha_{(\sbsigma)})\ ,
\end{equation}
and finds
\begin{equation}\label{f-sep-general}
f^{(\sbsigma)}_\sbalpha = \frac{1}{2^{K-|\sbsigma|}}\,
\prod_{i=1}^K\, u_i\ ,
\end{equation}
where \begin{equation}
  u_i =  \left\{ \begin{array}{ll} 1 +
(-1)^{\alpha_i}\, a_i  \ , & \ \ \sigma_i=0 \\  1 -[\alpha_i +
(-1)^{\alpha_i}\, b_i] \ ,& \ \ \sigma_i=1
\end{array} \right. \ ,
\end{equation}
with
\begin{equation}\label{}
    a_i = |\< \psi_i|\varphi_i\>|^2\ , \ \ \ b_i = \frac{a_i}{d}\
    .
\end{equation}
Hence, a $\bsigma$--invariant state $\mathcal{I}^{(\sbsigma)}_{\bf
f}$ is multi-separable iff
\begin{equation}\label{Kf||}
    f^{(\sbsigma)}_{\sbalpha } \leq \frac{1}{2^{|\sbalpha|}}
    \left(\frac{2}{d}\right)^{\!|\sbsigma\sbalpha|}\ ,
\end{equation}
where
$\bsigma\!\!\balpha=(\sigma_1\alpha_1,\ldots,\sigma_K\alpha_K)$,
and
\begin{equation}\label{ff}
f^{(\sbsigma)}_{\sbalpha } \leq  f^{(\sbsigma)}_{\sbbeta }\ , \ \
\ \mbox{for} \ \ \ |\!\!\balpha|> |\!\!\bbeta|\ .
\end{equation}
In particular for $2K$-partite Werner state, i.e.
$\bsigma=(0,\ldots,0)$ one has
\begin{equation}\label{}
    q_{\sbalpha } \leq \frac{1}{2^{|\sbalpha|}}\ ,
\end{equation}
whereas for $2K$-partite isotropic state, i.e.
$\bsigma=(1,\ldots,1)$, one finds
\begin{equation}\label{}
    p_{\sbalpha } \leq \frac{1}{d^{|\sbalpha|}}\ .
\end{equation}

Finally, one may prove that  a general $2K$--partite
$\bmu$-invariant state is $2K$-separable iff it is $\bnu$--PPT for
all binary vectors $\bnu$ and it is $A \ot B$ bi-separable iff it
is $(1\ldots 1)$--PPT.

\subsection{Reductions}

It is evident that reducing the $2K$ partite $\bsigma$--invariant
state with respect to $A_i \ot B_i$ pair one obtains
$2(K-1)$--partite $\bsigma_{(i)}$--invariant state with
\begin{equation}
\bsigma_{(i)} =
(\sigma_1,\ldots,\check{\sigma}_i,\ldots,\sigma_K)\ ,
\end{equation}
where $\check{\sigma}_i$ denotes the omitting of $\sigma_i$. The
reduced state lives in
\begin{equation}\label{}
    \mathcal{H}_1 \ot \ldots \check{\mathcal{H}}_i \ot \ldots \ot
    \check{\mathcal{H}}_{i+K} \ot \ldots \ot \mathcal{H}_{2K}\ .
\end{equation}
The corresponding fidelities are given by
\begin{equation}\label{}
    f^{(\sbsigma_{(i)})}_{(\alpha_1\ldots\alpha_{K-1})} =
    \sum_{\beta}\,
    f^{(\sbsigma)}_{(\alpha_1\ldots\alpha_{i-1}\beta\alpha_i\ldots\alpha_{K-1})}\
    .
\end{equation}
Note, that reduction with respect to a `mixed' pair, say $A_i \ot
B_j$ with $i\neq j$, is equivalent to two `natural' reductions
with respect to $A_i \ot B_i$ and $A_j \ot B_j$ and hence it gives
rise to $2(K-2)$--partite invariant state. This procedure
establishes a natural hierarchy of multipartite invariant states.

\section*{Appendix}
\def\theequation{A.\arabic{equation}}
\setcounter{equation}{0}

The 4-partite Werner state $\mathcal{W}^{(2)}_{\bf q}$ is $12|34$
(or $A \ot B$) separable iff there exists a bi-separable state
$\varrho$ such that $\mathcal{W}^{(2)}_{\bf q} =
\mathcal{D}^{(2)}\varrho$. Consider an extremal $A|B$ separable
state $\varrho = P_A \ot P_B$ where $P_A$ and $P_B$ are bipartite
projectors living in $\mathcal{H}_A =\mathcal{H}_B= \mathcal{H}_1
\ot \mathcal{H}_2 \equiv (\mathbb{C}^d)^{\ot 2}$. Simple
calculations give rise to the corresponding fidelities $q_\sbalpha
= \mbox{Tr}(\varrho\, \mathbf{Q}^\sbalpha)$:

\begin{widetext}
\begin{eqnarray}\label{qqqqA}
q_{00} &=& \frac 14\, \Big\{ 1 +\mbox{Tr}_2\left( \mbox{Tr}_1 P_A
\cdot \mbox{Tr}_1 P_B\right) + \mbox{Tr}_1\left( \mbox{Tr}_2 P_A
\cdot
\mbox{Tr}_2 P_B\right) + \mbox{Tr}_{12} (P_A\cdot P_B)\Big\} \ , \nonumber\\
q_{01} &=& \frac 14\, \Big\{ 1- \mbox{Tr}_2\left( \mbox{Tr}_1 P_A
\cdot \mbox{Tr}_1 P_B\right) + \mbox{Tr}_1\left( \mbox{Tr}_2 P_A
\cdot
\mbox{Tr}_2 P_B\right) - \mbox{Tr}_{12} (P_A \cdot P_B)\Big\} \ ,\nonumber\\
q_{10} &=& \frac 14\, \Big\{ 1 + \mbox{Tr}_2\left( \mbox{Tr}_1 P_A
\cdot \mbox{Tr}_1 P_B\right) - \mbox{Tr}_1\left( \mbox{Tr}_2 P_A
\cdot
\mbox{Tr}_2 P_B\right) - \mbox{Tr}_{12} (P_A\cdot P_B)\Big\} \ ,\\
q_{11} &=& \frac 14\, \Big\{ 1 - \mbox{Tr}_2\left( \mbox{Tr}_1 P_A
\cdot \mbox{Tr}_1 P_B\right) - \mbox{Tr}_1\left( \mbox{Tr}_2 P_A
\cdot \mbox{Tr}_2 P_B\right) + \mbox{Tr}_{12} (P_A\cdot P_B)
\Big\} \ ,\nonumber
\end{eqnarray}
\end{widetext} where $\mbox{Tr}_1$ denotes a partial trace in
$\mathcal{H}_1 \ot \mathcal{H}_2$. Therefore, for a general $A|B$
separable state (convex hull of extremal product states) one
obtains from (\ref{qqqqA}):
\begin{equation}\label{}
    q_{01},q_{10},q_{11} \leq q_{00}\ ,
\end{equation}
and
\begin{equation}\label{A1/2}
    q_{01} + q_{10}  \leq \frac    12\ .
\end{equation}
Note, that above conditions are equivalent to the condition
(\ref{11PPT}) for $(11)$--PPT. The third equation in (\ref{11PPT})
implies (\ref{A1/2}) whereas the first (second) and third gives
$q_{00} \geq q_{01}$ ($q_{00} \geq q_{10}$). Note, that
4-separable Werner state is necessarily bi-separable but the
converse is not true. Taking $\varrho = P_A \ot P_B$ such that
\begin{equation}\label{}
    \mbox{Tr}_2\left( \mbox{Tr}_1 P_A
\cdot \mbox{Tr}_1 P_B\right) = \mbox{Tr}_1\left( \mbox{Tr}_2 P_A
\cdot \mbox{Tr}_2 P_B\right)\ ,
\end{equation}
and $\mbox{Tr}_{12} (P_A\cdot P_B) \neq 0$ one obtains a
bi-separable Werner state $\mathcal{D}^{(2)}(\varrho)$ with
\begin{equation}\label{}
    q_{01}=q_{10} < q_{11} \ ,
\end{equation}
which contradicts 4-separability.

\acknowledgments  This work was partially supported by the Polish
State Committee for Scientific Research Grant {\em Informatyka i
in\.zynieria kwantowa} No PBZ-Min-008/P03/03.

 \end{document}